\documentclass[twocolumn,showpacs,amssymb,aps]{revtex4}

\usepackage{graphicx}
\usepackage{psfig}
\usepackage{dcolumn}
\usepackage{bm}

\def\lessim{\lower.5ex\hbox{$\; \buildrel < \over \sim \;$}}
\begin{document} \hbadness=10000
\topmargin -1.cm

\title{Deconfinement  energy threshold: analysis of hadron yields 
at 11.6 \mathversion{bold}$A$ GeV}
\author{Jean Letessier$^a$, Johann Rafelski$^{a,b}$, and Giorgio Torrieri$^{b,c}$}
\affiliation{%
$^a$Laboratoire de Physique Th\'eorique et Hautes Energies\\
Universit\'e Paris 7, 2 place Jussieu, F--75251 Cedex 05;\\
$^b$Department of Physics, University of Arizona, Tucson, Arizona, 85721, USA;\\
$^c$Department of Physics, McGill University, Montreal, QC H3A-2T8, Canada}

\date{January 5, 2005}

\begin{abstract}
We analyze the hadron yields obtained at the AGS in the range 11--11.6 $A$ GeV 
and discuss strategies to identify possible deconfinement at this 
energy scale. These include consideration of chemical non-equilibrium  
at hadronization, and the study of (multi)strange hadrons. We find that 
the totality of experimental results available favors the 
interpretation  as hadron freeze-out  at the 
phase boundary  between confined and deconfined phase.

\end{abstract}

\maketitle
One of  the  most interesting issues, in the field of relativistic heavy ion collisions,
is the understanding of the thresholds in reaction  energy, and system size, beyond
which  the formation of the color deconfined  partonic state occurs dominantly. 
In the central collisions of Au--Au ions at BNL--RHIC, at the 
top energy range ($\sqrt{s_{\rm NN}}=200$ GeV), it is 
generally believed that a color deconfined state has been formed \cite{Part}.
The  strange antibaryon production systematics
lead us to  believe that this is also the case at the top energy in 
central Pb--Pb ion reactions at $\sqrt{s_{\rm NN}}=17.2$ GeV 
(158 $A$ GeV/c Pb beam colliding with fixed target)  
at CERN--SPS \cite{torrieri_njp}. 
Experimental exploration of lower energy collisions at SPS 
down to $\sqrt{s_{\rm NN}}=6.3$ GeV 
(20 $A$ GeV/c) suggests a possible change in strangeness 
production pattern \cite{Gaz,Raf03}.
In this analysis, we address the  experimental 
program at BNL--AGS at $\sqrt{s_{\rm NN}}=4.8$ GeV (11.6 $A$ GeV/c).
Our objective is to apply the  methods we used in the study of 
the SPS and RHIC data in order to identify similarities and  differences 
in the  hadron production pattern.

The tool used in this study of hadron production is the 
Statistical Hadronization (SH) model  introduced by Fermi in 1950~\cite{Fer50}. 
In  50 years, SH has matured to a full
fledged tool, in the  study of soft strongly interacting particle 
production,  capable to describe   in detail hadron  abundances 
once the statistical grand canonical method and 
 the full spectrum of resonances is included  \cite{Hag65}.
The key SH  parameters of interest 
are the temperature $T$ and  the baryochemical potential $\mu_{\rm B}$. 
It is generally accepted that  as the energy of the colliding 
nuclei varies, the 
properties of the hadronic gas (HG) and quark--gluon plasma phases (QGP)
are explored in a wide domain  of $T$ and  $\mu_{\rm B}$ , 
see Fig.\,2.~in~\cite{Hag80}.

Two  values of temperature can be determined  in  statistical 
hadronization studies:  chemical freeze-out $T_{\rm ch}$,
 which determines  particle abundances, 
and a thermal freeze-out $T_{\rm th}$, defined by 
the condition where the momentum 
spectra stop evolving, {\it i.e.}, particles stop 
interacting elastically.  We analyze here solely the 
yields of hadronic and in particular strange particles
 and, hence from now, on we imply  $T=T_{\rm ch}$. 
For $T$ to be uniquely determined for all particles considered 
the chemical freeze-out must  occur rather fast. This requires that 
particles stop interacting just  after formation (hadronization). 
 In fact, the same freeze-out condition 
 explains, at  high energy SPS \cite{torrieri_njp}, 
and at  RHIC \cite{florkowski},  both particle
abundances and spectra. 
This is the so called single freeze-out model expected to apply
in presence of sudden 
hadronization of a rapidly expanding supercooled 
quark--gluon fireball \cite{sudden}. 

The single freeze-out model, allowing for chemical non-equilibrium,
 is consistent with the 
observation by invariant mass method of abundantly produced 
hadron resonances by, {\it e.g.}, the STAR and NA49
collaborations \cite{Markert,fachini}. 
This results demonstrates  that resonance 
decay products in essence have not scattered before 
their observation \cite{TorrieriRes}. 
Hence  resonance  decay spectra
 contribute  to the stable particle spectra, altering the purely thermal
shape. An analysis of this modified shape   yields
 the  universal freeze-out condition \cite{torrieri_njp,florkowski}. 
On the other hand,   studies 
in which resonance decays are for simplicity ignored have as result a
 different thermal  freeze-out condition  for practically each particle
 considered, which differs from the chemical freeze-out condition 
obtained using hadron yields. These may serve  as a
comparison benchmark but cannot be used otherwise. 

The question of chemical equilibration arises in the 
context of chemical freeze-out studies. The introduction,
into the data analysis, of the strangeness phase space occupancy
parameter $\gamma_s$ was motivated by the recognition that 
strangeness (or equivalently strange hadron)
build-up in microscopic reactions will rarely lead to abundance  expected in  
chemical equilibrium,  $\gamma_s=1$  \cite{Raf91}.
The parameter $\gamma_s$  characterize, independently of 
temperature, the particle--antiparticle pair yield.
 Subsequently, the light quark phase space 
occupancy   $\gamma_q$ was introduced \cite{gammaq}. This allowed
to describe  particle yields for the 
case  of fast hadronization. Namely, the phase
space density of  quarks in  QGP is very different from 
that obtained evaluating the yields of valance quark content in final
state hadrons.  

Thus, even if during its temporal
evolution a QGP fireball converges to chemical equilibrium, 
the final hadron state generated on the scale of a few fm/c 
should emerge  showing a  pattern
of subtle  deviations  from   chemical equilibrium which 
contain interesting information about the physics we are
exploring \cite{Raf01kc,CUP}.
The claim  that, in general,  overall chemical 
equilibrium is reached in relativistic heavy ion collision is 
 based on an analysis that presupposes this result \cite{PBM}. 
In this paper, we show that while chemical  equilibrium hypothesis for 
the 11 $A$ GeV reactions can be considered, the experimental results 
available today favor chemical under saturation of both light 
and strange quark 
phase spaces, with both $\gamma_s,\gamma_q <1$.

Let us first explain how the different parameters of 
the SH model can be determined using the experimental data in a step
by step process, rather than in 
 a global fit, which erases much of the physics insight.
We will show that it is
 possible to describe a partial subset of data with a partial subset
of parameters  since certain types of particle ratios 
  are sensitive to subsets of statistical parameters. Of particular 
importance in this discussion is that  the chemical non-equilibrium
parameters $\gamma_s, \gamma_q$, require to be determined 
ratios of certain rarely produced particles.
 
a) The chemical fugacities $\lambda_q$, $\lambda_s$, see~\cite{Raf91}, and $\lambda_I$
(see e.g Eq.\,(\ref{pirat} below)  
  can be  determined  rather  precisely and  independent of other statistical parameters
  considering  the anti-particle to particle ratios.  There are many such ratios 
available allowing to fix the three parameters, check consistency and  to make ratio predictions
based on a partial data set. Success of this part of SH model analysis has no predictive power 
regarding  the value of the remaining 4 statistical parameters. Alas, we see again and again
the argument to the contrary, in particular some workers claim based on the success 
of describing particle to antiparticle ratios that  there is chemical equilibrium,
$\gamma_s=\gamma_q=1$, which is plainly wrong.  

b) If one considers from that point on only the product of particle with antiparticle
  yield, to a very good approximation one obtains reduced  yield ratios which are independent of
  $\lambda_q$, $\lambda_s$ and $\lambda_I$ . Than
to determine the  ratio $\gamma_s/\gamma_q$ (or equivalently $\gamma_s$ when
$\gamma_q=1$) we consider   ratios of particles with unequal strangeness
content, e.g. $\phi/\sqrt{{\rm K}^+{\rm K}^-}\propto \gamma_s/\gamma_q f_1(T)$
In such a ratio generally there  is  a   correlation between $\gamma_s/\gamma_q$ and $T$
  and thus at least  two  such  ratios are required to
determine $\gamma_s/\gamma_q$ and $T$, see for example Ref.\cite{Rafelski:2002ga}, figure 4\,.

c) Similarly, comparing yields of particles with different quark number content, e.g.
mesons with baryons one identifies the value of  $\gamma_q$, as noted in
{\it e.g.}, Section IVD in Ref.\,\cite{becattini}. Consider here as  example
$\Lambda \overline\Lambda/{\rm K}^+{\rm K}^-\propto \gamma_q f_2(T)$; we see that
at least a second  such ratio is required, e.g.  
$\sqrt{\Xi^-\overline\Xi^+}/\phi$ or/and $N\overline N/\pi^+\pi^-$ etc. to determine 
both $\gamma_q $ and $T$ 
or/and  we take  $T$ from the study of  $\gamma_s/\gamma_q$ and $T$ above. We expressly
did not introduce above   pion yields as these derive from resonance decays including heavy 
baryons, and thus their yield is not strictly the yield of mesons. Moreover
the pion yields are theoretically most uncertain considering 
possible extension of the  hadron mass spectrum to higher mass, and uncertainties 
about number of pions produced in cascading decays of heavy resonances. In absence 
of experimental data regarding multistrange particles there is little choice but to 
compare the yields of pions to  nucleons in the analysis of $\gamma_q $ and $T$.
 
d) Each individual total particle yield is  proportional to  the volume parameter  $V$ aside
of a strong dependence on $T$, and a lesser dependence on   all the other SH parameters. This
allows to fix $V$ and test the consistency of the individual findings about
the other parameters made above. 

In a more modern approach one makes a global and simultaneous fit of all parameters
minimizing the global error ($\chi2$). This
   can only work if there is sufficiently wide scope of experimental data as required in
the individual steps above. Practical experience shows that an incomplete data set
combined with the presence of a significant measurement error  
   makes the task of unfolding $\gamma_s, \gamma_q$ and $T$   difficult. 
In this environment   some workers think that it is better to assume  
 $\gamma_q= 1$,  we will discuss this at length below. Here we note
that  the predictive power of SH model  can be tested:
we can create choosing a set of SH model parameters a set of  particle yields,
 and  add in random errors. If the requirements on availability of measurements
  identified in a)--d) are respected in the data set considered,
one can   fit such an  artificial data set to obtain  the
  `creating' set of statistical parameters. However, when large yield errors 
are introduced, and key particles eliminated, 
the  expected  solution will not always be found   in the likelihood analysis. 

Considering such  studies   it is important    to remember  that   the SH model is not being
verified or tested,  its validity can be seen in its predictive power which encompasses
particle  yields that vary by typically four  orders of magnitude,   
 varying between $\cal{O}(0.1)$ and
$\cal{O}(100)$. The issue before us is, {\it  if and when} we can use statistical
significance  of the fit to infer the values of the parameters, and the insights
  obtained in such   model studies are affirming this.  For this reason it makes
sense to look for most likely solutions using a full parameter set motivated by
theoretical considerations.

The parameter  $\gamma_q$ has been introduced into SH model not as a
fit parameter, but in consequence of a theoretical development. 
It is the  quantity which is expected to differ from unity considering hadron
freeze-out from   a rapidly expanding system
and/or undergoing a sharp phase transformation.  In such
consideration $\gamma_q=1$ only when
there is a relatively long time available for  chemical re-equilibration. 
Moreover,  a large value of
$\gamma_q$ allows the hadron gas phase to absorb without volume increase the
enhanced entropy content of the deconfined phase in which color bonds are broken.
The ability of the two models ($\gamma_q=1$ and $\gamma_q \ne 1$) to describe
data should be compared in a quantitative way in the study of
statistical significance. Evidence for $\gamma_q > 1$
constitutes evidence for a rapidly evolving system emerging from a
deconfined phase~\cite{Letessier:1998sz,Rafelski:1999xu}.

Statistical significance \cite{pdg}, 
defined as the probability of the data fit to a model to be
  of the obtained quality, provided that the
 model under consideration is `true', and errors are purely
due to experiment, is an appropriate tool for the comparison 
of model variants. This approach
takes into account both the number of fit parameters 
and the `goodness' ($\chi2$) of the fit.
In the absence of the experimental data necessary 
for strict falsification, it has been 
argued by others that a comparison of statistical significance 
is a more sound procedure than turning to
a more restricted and less general model with the 
fewest number of fit parameters~\cite{philosophy}.
Specifically in our context, if the introduction of 
$\gamma_q$ as a fit parameter yields results expected
from theoretical considerations, and at the same time 
this step   raises statistical significance
considerably, and furthermore makes the behavior
of best fit parameters more consistent with expectation
 once experimental conditions  are varied
(collision energy and centrality \cite{centrality}), than it is appropriate
to conclude that current  experimental data favor  $\gamma_q \ne 1$.

The methods of SH analysis are described elsewhere in great detail. 
We refer the reader to SHARE (statistical hadronization with resonances), 
the public SH suit of (FORTRAN) programs
 which we use in this analysis \cite{share}. 
The parameter set of SHARE 
comprises  also the 
fugacity $\lambda_{\rm I}$, which describes the asymmetry 
in the yield of up  $u$ and down  $d$ valance quarks. This parameters
cannot be omitted (tacitly set to unity)   in the AGS physics environment,
since the initial state isospin asymmetry is not diluted by a very large 
produced particle yield. 
 To understand the relevance of $\lambda_{\rm I}$, note that,  in Boltzmann
approximation prior to resonance decays (subscript `o'), we have for the
pion ratio:
\begin{equation}\label{pirat}
{\pi_{\rm o}^+\over \pi_{\rm o}^-}=\lambda_{\rm I}^2\equiv(\lambda_u/\lambda_d)^2.
\end{equation}
The reader will notice that  we abbreviate particle  yields, 
{\it e.g.}, $Y_{\pi^\pm}$ by particle name $\pi^\pm$. 

Using SHARE,  we evaluate for a set of 
statistical parameters the particle yields and as appropriate, ratios, 
 and then find in least square minimization process 
the best parameter for the 
experimental results applicable to the considered  collision system.
We use here 
the most central event trigger available  (5\%). 
For the top energy at  AGS, 
we adopt nearly the totality of 
the experimental results considered  by prior authors as stated 
in Ref.\,\cite{becattini}, listed here and comprising 
the following further developments: 

i) 
 Another way
to introduce   fugacity $\lambda_{\rm I}$ is to conserve electric 
charge, see \cite{becattini}.  Thus   $d>u$ asymmetry 
is equivalent  to the requirement  that 
the total charge $Q$ of the hadronic fireball is a fixed 
fraction $f=0.391$ of  the total
baryon number $B-\overline{B}=363\pm10$ (1) ({\it we point out and 
count the number of measurements by having behind each value used 
in parenthesis a sequential number}),   established 
by proton and neutron content of 
the colliding gold ions, {\it i.e.}, $Q=142\pm 5$~(2). 
We note that these values arise from our choice of the centrality
trigger condition see, {\it e.g.}, table I, in Ref.\,\cite{Ahle60}.

ii) 
We adopt a slightly different strategy in use the available particle 
yield data, in
order to limit the influence of the systematic error on the fit result: 
rather than to fit $K^-=3.76\pm0.47$ yield, 
we fit $\mbox{K}^+/\mbox{K}^-=6.32\pm0.65$~(3)
which is seen in Fig.\,6 of Ref.\,\cite{Ahle60}.  

iii) 
We explore what happens when we include in the 
 fit the (redundant) but independently measured precise ratio 
$\mbox{K}^+/\pi^+=0.197\pm0.013$~(4) (value taken 
from conclusions to  Ref.\cite{Kslope}) , 
which compares  to the implied ratio 
$\mbox{K}^+/\pi^+=0.177\pm 0.037$ derived from the 
individual yields we use:  
$\mbox{K}^+=23.7\pm2.86$~(5) (from table I in   Ref.\,\cite{Ahle60},
including systematic error),
 and $\pi^+=133.7\pm 9.93$~(6) (from extrapolation of  Ref.\,\cite{BecRed} 
as used in \cite{becattini})
The results we present are obtained 
including this as an independent measurement.
Removing this ratio yields a reduction of 
P-value from, {\it e.g.}, 65\% to 55\%
but does not affect materially the results we discuss.

iv) 
The other particle yield or, respectively, ratio values we use are:
 $\Lambda=18.1\pm1.9$ (7) , 
$\overline\Lambda=0.017\pm0.005$ (8), 
and $p/\pi^+=1.23\pm0.13$ (9), in all of
these  following the procedure of Ref.\cite{becattini}. One 
of the reasons we did not deviate wherever possible 
from the data set of  Ref.\cite{becattini} was to assure 
that the results we obtain can be compared 
directly with earlier work.

v)
We further study the influence of the  recently published  E917  ratio
$\phi/\mbox{K}^+=0.03\pm 0.006$ \cite{phi}, obtained within and interval of 
$\delta y=\pm 0.4$ units of rapidity around the mid-rapidity. We choose from the 
other measurements presented ($\phi$ yield, $\phi/\pi$) this result as it is 
the most precise one, and suitable to a qualitative extrapolation 
to a total yield ratio.
We assume that the Gaussian shape rapidity distribution has
 $\sigma_\phi=0.8$ which is consistent with the data shown by E917 \cite{phi},
and the energy dependence systematics of  $\sigma_\phi$ obtained 
by the NA49 collaboration, see Fig.\,3 in \cite{phiNA49Nantes}. Since the
interval of $|\delta y|=0.4<\sigma_\phi=0.8$ and 
$\sigma_\phi=0.8\lessim {\sigma_{K^+}}=0.96\pm0.6$ ($\sigma_{K^+}$ is
from Ref.\,\cite{Ahle60}), to   \%-precision,
the relation of yields within the rapidity interval to the $4\pi$ total yield is
\begin{equation}
{\phi \over {\rm K}^+}|^{\!\!|}_{{\!\!|}_{4\pi}}\simeq 
        {\sigma_\phi\over \sigma_{\rm K}}\,
  {\phi \over {\rm K}^+}|^{\!\!|}_{{\!\!|}_{y\in (-0.4,+0.4)}} .
\end{equation}
Thus, we adopt  as the $4\pi$ ratio 
 $\phi/\mbox{K}^+\vert_{4\pi}=0.025\pm 0.006$ (10), which
expresses the fact that the $\phi$ is expected to have a narrower 
rapidity distribution than the ${\rm K}^+$.

The SH model comprises  aside of the already mentioned
5 parameters  $T$, $\mu_{\rm B}$
(equivalently valance light quark fugacity $\lambda_q$), $\gamma_s$,
$\gamma_q$, $\lambda_I$ also the  fireball volume $V$ and $\lambda_s$,
the strange quark fugacity. In principle  this last parameter can be fixed by 
the requirement that the strangeness and anti-strangeness content 
balances in the fireball, {\it i.e.}, within the grand-canonical ensemble, 
\begin{equation}\label{scons}
\langle s \rangle = \langle \bar s \rangle .
\end{equation}
However, we hesitate to use this constraint since:

a) When chemical analysis includes both $\gamma_s,\gamma_q$ and one of 
the sides of above equation 
involves in essence just one particle species (K$^+$ here)
this constraint  has in general several possible solutions. 

b) In the condition  Eq.\,(\ref{scons}), 
all the other measurements combine predicting the yield  of K$^+$
required by the strangeness balance. This  results in a consistency test
of the data sets obtained from several experiments employing different 
instruments.

 All this speaks against implementing this constraint 
and the  results we present were obtained not enforcing   Eq.\,(\ref{scons}).
We will show by how much this constraint could be violated (in relative terms 
about 10\%) and  also show that its enforcement is not in essence 
altering the results shown.

\begin{figure}[tb]
\resizebox{3.in}{!}{\rotatebox{0}{\includegraphics{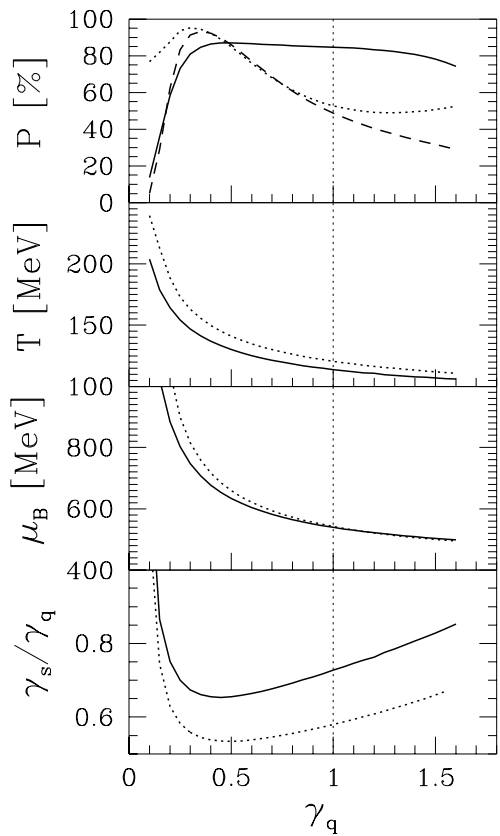}}}
\caption{\label{AGSfit} 
From top to bottom: statistical significance of the fit for 
the data set, see text, at a fixed value of $\gamma_q$; chemical freeze-out
temperature $T$; baryochemical potential $\mu_{\rm B}$; strangeness phase space 
occupancy divided by light quark occupancy $\gamma_s/\gamma_q$, all as function of 
the prescribed to the fit
 light quark phase space  occupancy $\gamma_q$. Dashed line in upper section
includes in fit the $\phi$. Dotted lines are obtained imposing strangeness 
conservation. Vertical line, at $\gamma_q=1$, is for orientation only.
}
\end{figure}

In this study,  we have 
7 free parameters and up to 10  data points, of which one comprises partial
 redundancy. We will validate  the fit instead
of $\chi^2$ by the associated significance level  $P[\%]$ which is obtained
in SHARE using the CERN library ``PROB'' procedure \cite{cernlib}.  $P[\%]$
is a function of the number of parameters $p$, and measurement points $r$. 
When these numbers are small, and in particular, 
the number of measurements is not much greater than the number of parameters,
the value of total  $\chi^2$ must be much smaller than what is usually expected
for the  90\% significance level result ($\chi^2\simeq p-r$). Using this 
tool we find that `good looking' figures arise for
 $P\simeq 15\%$ applicable to the result shown in Ref.\,\cite{becattini}. Such
a low significance  level could be result of chance, but more likely it
means that either  the error on some of the 
 measurements considered is in realty bigger,
or the theoretical  model needs further refinement, 
such as, {\it e.g.}, light quark 
chemical non-equilibrium.

In view of the above, we first wish to 
understand if the $\gamma_q=1$ choice (light quark chemical
 equilibrium) is  compelling. Using the SHARE1.2 package 
(without particle widths), we obtained the significance 
level ($P$-value) of our fits which we show 
in the top section of  Fig.~\ref{AGSfit}. We see (solid line)
the significance level  of our SHARE fit with 9 first 
experimental results, which has a plateau near 80\% in the range
$0.3<\gamma_q<1.5$. Eliminating the one redundant data point
the significance  level drops to 55\% and the range of acceptable 
values of $\gamma_q$ widens further. The  vertical 
  line,  in Fig.~\ref{AGSfit}, is placed at the chemical equilibrium,
with  assumed value $\gamma_q=1$.

The dashed line, in the top panel 
of Fig.~\ref{AGSfit},  represents the significance  level 
arising when the new 10th experimental point $\phi/\mbox{K}^+$ 
is included in the analysis. 
The best fit is pushed well below chemical equilibrium with
$\gamma_q\simeq 0.4$,  
and the significance level rises nearly to 90\%. 
 As this result shows, there is 
no compelling reason  to consider only the case of
 light quark chemical equilibrium. Indeed this added experimental
result favors chemical non-equilibrium for light quarks.
The full chemical equilibrium $\gamma_s=1,\gamma_q=1$ 
appears inconsistent with the data set as can be 
seen inspecting the bottom panel of  Fig.~\ref{AGSfit} where the 
ratio  $\gamma_s/\gamma_q$ is presented. 

 Dotted  lines, in  Fig.\,\ref{AGSfit}, were obtained enforcing strangeness
conservation condition, Eq.\,(\ref{scons}),
as was done in  Ref.\,\cite{becattini}.  This condition does not
influence in essence the discussion we present though in detail 
minor differences arise. The two strategies we pursued in finding 
the best fits assure us that we have obtained the 
best fit in a situation involving many parameters, 
 where several local minimal points are present.
Our analysis shows that the suggestion 
that at low reaction energies the 
freeze-out of hadrons occurs well below the theoretical phase boundary 
between QGP and HG is not fully justified, as it is result of the 
assumption of chemical equilibrium. 

Specifically, we find  for the best fit with $\phi/\mbox{K}^+$ 
a hadronization temperature $T=142\pm3$  MeV and the baryochemical potential 
$\mu_{\rm B}=708\pm 60$ MeV. 
We realize that  both these values  $T$ and $\mu_{\rm B}$ 
appear at first sight to be beyond  the  phase boundary 
region of the hadron  gas, in the  deconfined  domain, given the 
estimate for the  critical  boundary \cite{Lattice}, and the 
related development of the liquid QGP phase model \cite{liquid}. However, one 
has to realize that the relatively small values for $\gamma_q$ and $ \gamma_s$ we
obtained imply a significant reduction in the equilibrium 
QGP pressure, which pushes up the  temperature of the 
non-equilibrium phase boundary. This opens the 
question if indeed the hadron yield results seen at AGS are not the product
of the breakup of a deconfined state. In our earlier study of light
ion collisions at AGS, we had noticed that the results allowed 
interpretation in terms of both confined and deconfined baryon rich fireball 
formation \cite{Let94}.  

\begin{figure}[tb]
 \resizebox{3.in}{!}{\rotatebox{0}{\includegraphics{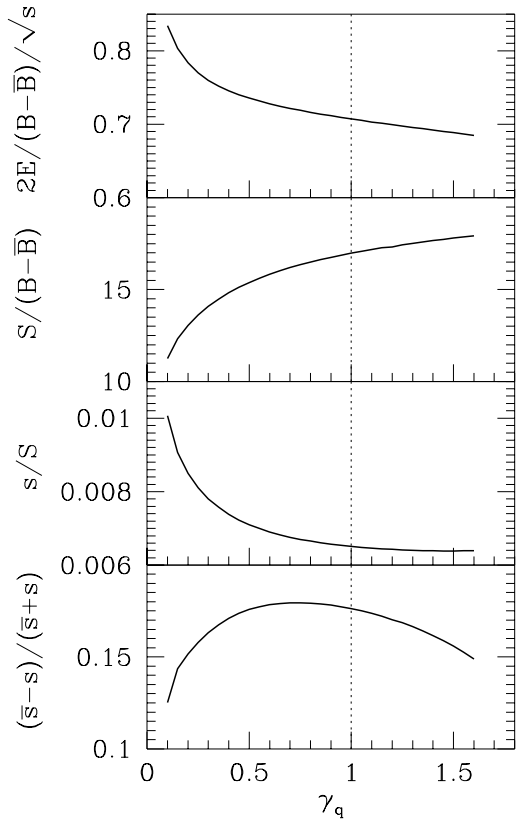}}}
\caption{\label{AGSprop} 
Physical properties of the thermal fireball at chemical breakup as function
of $\gamma_q$.
 From top to bottom: thermal energy content weighted
 with $\sqrt{s_{\rm NN}}=2.4$ GeV; entropy per baryon $S/(B-\bar B)$,
 strangeness per baryon $s/(B-\bar B)$ and
the strangeness asymmetry of the fit,
$(\bar s-  s)/(\bar s+  s)$.
}
\end{figure}

 We can only speculate about the physical reasons behind these
small values of favored phase space occupancies: $\gamma_q=0.35\pm0.27$,
 $\gamma_s=0.23\pm0.18$. For example, if the deconfined state were to be 
well characterized at high baryon density 
by effective heavy quarks, with $m \simeq 350$ MeV, these quark equilibrium 
yields after  hadronization at $T=142$ MeV
 could well appear much below chemical equilibrium.
 This is consistent with the low value of entropy per baryon, Fig.~\ref{AGSprop}, 
$S/(B-\bar B) =14\pm2$. This situation is opposite to what 
happens for the  small baryon density case (RHIC), 
with effectively small quark 
masses at hadronization,  where the hadronization 
process overpopulates the 
hadron yield.

Considering together  the AGS, SPS and  RHIC  energy ranges, 
we recognize that the   hadronization temperature 
arises from a combination of several effects: the location of the equilibrium
critical curve in the $T,\mu_{\rm B}$-plane, shift in the critical curve due
to chemical non-equilibrium conditions, and 
supercooling due to dynamics of the expanding fireball. The last effect is 
particularly significant at RHIC while the first is probably most
relevant at AGS as we noted above.   At RHIC,
the fast transverse expansion is seen in many observables,
and it is generally believed that a new partonic phase has
been created. For an equilibrium system the  
 phase cross-over would be expected at $T=164\pm10$ MeV \cite{Lattice}.
However, the wind of colored particles expanding against the color-non-conducting
vacuum can go on until the  sudden breakup  near to $T=140$\,MeV \cite{sudden,CUP}.
 Aside of theoretical arguments, and 
fit results,  such low hadronization temperatures are consistent 
with the   observed resonance yield,  for 
$K^*, \phi$ \cite{RafRHIC}, as well as $\rho^0, f_0$ and $ \Lambda(1520) $  \cite{Torinprep}.

Returning to the discussion of the 
physical properties of the fireball:
 aside of entropy per baryon  mentioned above, we obtain other 
 relevant physical properties of the source of produced hadrons
 using the statistical parameters to 
evaluate the phase space properties. We present, in Fig.~\ref{AGSprop}  
from top to bottom, the thermal energy per baryons as fraction of the 
center of momentum energy available in the reaction,  the single 
particle entropy per baryon, strangeness per baryon and the
strangeness asymmetry in the fits we consider when strangeness 
conservation  is not enforced.  For the interesting range
$\gamma_q\simeq 0.4$,  the value  $s/S\simeq 0.008\pm0.001 $ is significantly below
the SPS and RHIC level \cite{Raf03}. This is consistent with the notion that 
strangeness yield rises with energy faster than light quark yield, as 
can be argued considering the  mass--energy threshold for the production
of these flavors, and the time available in the collision.

\begin{figure}[bt]
\resizebox{3.in}{!}{\rotatebox{0}{\includegraphics{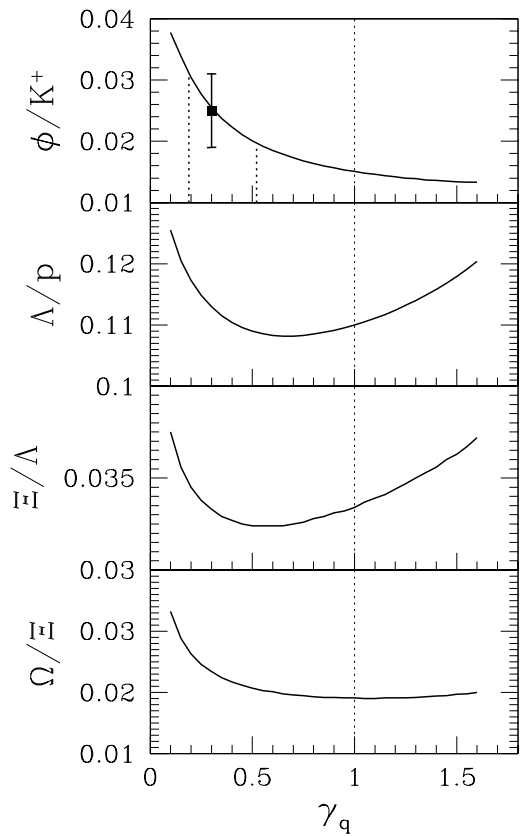}}}
\caption{\label{AGSPhiHyp} 
 From top to bottom:  relative yields of 
 $\phi/\mbox{K}^+$, $\Lambda/p$, $\Xi^-/\Lambda$ and $\Omega/\Xi^-$
  as function of $\gamma_q$. The  experimental yield  $\phi/\mbox{K}^+$ is
placed at best value of $\gamma_q$ and the range (1\,s.d.) of 
possible $\gamma_q$ is indicated by vertical dots.
}
\end{figure}

\begin{figure}[tb]
\resizebox{3.in}{!}{\rotatebox{0}{\includegraphics{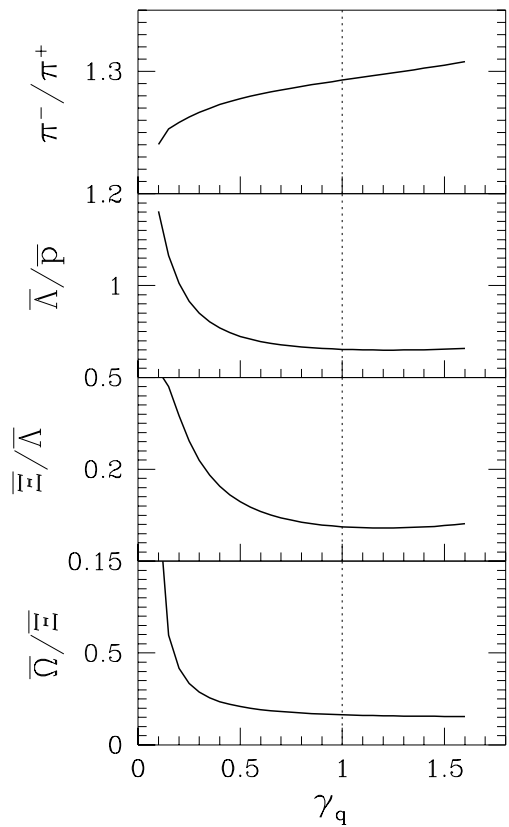}}}
\caption{\label{AGSPhiAntiHyp} 
From top to bottom:  relative yields of $\pi^-/\pi^+$,  
$\overline{\Lambda}/\bar p$,  $\overline{\Xi^-}/\overline{\Lambda}$ and
$\overline{\Omega}/\overline{\Xi^-}$,  as function of $\gamma_q$. 
}
\end{figure}

Considering that one measurement ($\phi/\mbox{K}^+$) pushes the discussion of AGS 
results  toward a  breakup of a deconfined state, we wish to understand  if
other measurements could confirm and solidify this result.
We present, in Figs.~\ref{AGSPhiHyp} and \ref{AGSPhiAntiHyp}, 
the variation  of  hadron ratios as function of $\gamma_q$ with 
other relevant statistical parameters varying as indicated in Fig.~\ref{AGSfit}.
We see  considerable sensitivity of hadron ratios considered 
to  the value of $\gamma_q$. We place the $\phi/\mbox{K}^+$ experimental
result in the  top panel of Fig.~\ref{AGSPhiHyp} at an appropriate value of 
$\gamma_q$,  indicating with dotted lines 
the range of  $\gamma_q$ consistent with the measurement error.
 Availability of  other strange hadron ratios 
would provide the consistency check required to confirm that  $\gamma_q<1$. 

The results, presented in Figs.~\ref{AGSPhiHyp} and \ref{AGSPhiAntiHyp}, 
suggest which of the experimental ratios are of relevance in the 
study of  chemical non-equilibrium. We see that some vary strongly,
and that   others vary little, {\it  e.g.}, $\Lambda/p$ ratio is flat to within 
10\% in the entire  range of $\gamma_q$ considered. Clearly, experiments 
will have great difficulty to reach  precision at \%-level to see
such variation.    We considered also 
baryon to meson ratios  involving $\Xi$, $\Lambda$  with  K (and antiparticles)
Mostly, these turned out  to be very insensitive, {\it i.e.}, flat to within 10\%. 
The exception are ratios of $\overline\Lambda/K$ which are flat only for 
$\gamma_q>0.4$. The variability for small $\gamma_q$ 
is accounted for in the ratios shown in
Fig.\,\ref{AGSPhiAntiHyp} involving $\overline\Lambda$.

We note that the ratio of $\overline{\Lambda}/\bar p\lessim 1$   
  at  $\gamma_q<0.2 $. The value of this ratio has been of 
considerable  interest \cite{antilambda}. 
The experimental result  $\overline{\Lambda}/\bar 
p=3.6^{+4.7+2.7}_{-1.8-1.1}$ favors a small value of $\gamma_q $
We further note that the ratio 
$\pi^-/\pi^+$, seen in Fig.\ref{AGSPhiAntiHyp}, is consistent with the
result  $\pi^-/\pi^+=1.23\pm0.02\pm15$--20\% obtained in central collisions 
\cite{E802pion}, and again a small value of  $\gamma_q $ is favored when comparing 
the experimental result to the SH model prediction, obtained fitting 
other experimental data.

This discussion confirms  the importance of multistrange hadrons and 
strange antibaryons as a valuable  hadronic signature. In this context, it
is important to understand the overall strangeness yield. Evaluating the 
number of valance strange quark pairs per baryon in terms of  the 
parameters characterizing the  phase space, we obtain for the  entire 
relevant range:
\begin{equation}\label{sbb}
{\bar s\over (B-\overline B)}=0.12\pm0.02\pm15\%, \ 0.2<\gamma_q<1.
\end{equation}
We show above the yield of $\bar s$ in emitted hadrons, since is greater than 
that of $s$ in our study. When we enforce strangeness conservation
Eq.\,(\ref{scons}), the presented result is reproduced.
The first  error stated in Eq.\,(\ref{sbb}) is the  variability of the
 yield as function of diverse  parameters. We see that it is well below the 
estimate of the propagation of the experimental errors into the 
theoretical yield, which we present  second  as the  estimated  error. 
 We note that  ratio Eq.\,(\ref{sbb}) 
includes   aside of open strangeness, also   a noticeable contribution 
contained in  $s\bar s$ mesons ($\phi$ and components in $\eta,\ \eta'$).

We see, in  Fig.~\ref{AGSprop}, that 
 there are more $\bar s$-hadrons  emitted   than $s$-hadrons.  
If this $\bar s> s$ asymmetry  were to be confirmed by more 
experimental data, this could be explained by
strangeness distillation phenomenon expected to 
occur at AGS energy \cite{strangelett}. After the excess of $\bar s$
is evaporated, the residue is a quark soup enriched with $s$-quarks,
a strangelett, which when metastable, may have escaped observation.

For central collisions, at the top AGS  energy, the transverse
slopes  of particle spectra  $T_\bot\simeq 200$ MeV   ({\it e.g.},
K-spectra, see Ref.\,\cite{Kslope}). 
This result leads to  the average radial  
expansion of the  AGS fireball  at 
$\langle v_{\rm r} \rangle=0.5$c \cite{Dobler:1999ju}.
Such a high
speed of radial expansion is, for us, hard to understand. On the 
other hand,  for a single freeze-out at $ T=143$ MeV we  report here, we estimate
the required $\langle v_{\rm r} \rangle=0.2$--0.3c (note that results of 
analysis \cite{Dobler:1999ju} do not fully apply as these were done assuming
chemical equilibrium yields of resonances).   The introduction of 
chemical non-equilibrium and the associated high chemical freeze-out 
temperature harbors  the potential for consistent  understanding of particle 
spectra and yields, also within a kinetic  reaction
model  \cite{Bratkovskaya:2004kv}.

In conclusion, we have presented a comprehensive and systematic
study of the  experimentally measured particle yield  ratios
obtained in  $\sqrt{s_{NN}}=4.8$  Au--Au collisions (11.6 GeV/c Au on
fixed target). We have analyzed the chemical freeze-out conditions
allowing the QGP associated chemical non-equilibrium. We have 
found that the present data sample is not sufficient to 
argue decisively for or against the QGP presence at AGS energy scale. 
However, consideration of the recently measured yields of $\phi$,
the $\pi^-/\pi^+$ along with indirectly evaluated $\overline\Lambda/\bar p$  
favors as the result of the  data analysis the 
chemical non-equilibrium 
hadronization at the phase boundary between the confined and 
deconfined baryon rich phase.  The results we presented 
indicate that exploration of the phase transition  between baryon rich 
confined and deconfined phases may be possible by means of
relativistic heavy ion beams in the energy range of  10 $A$  GeV/c.

\vspace*{.5cm}
LPTHE, Universit\'e Paris 6 et 7 is: Unit\'e mixte de Recherche du CNRS, UMR7589.
Work supported in part by a grant from: the U.S. Department of
Energy  DE-FG03-95ER40937 and DE-FG02-04ER41318, 
the Natural Sciences and Engineering research
council of Canada, the Fonds Nature et Technologies of Quebec. 
G. Torrieri thanks the Tomlinson foundation for support given.


\vskip 0.3cm

\end{document}